\documentclass[11pt]{article}
\usepackage{epsfig}
\def\one{1\hskip -.37em 1}     
\def\proj{E\hskip -.69em I}     
\begin{document}
\begin{titlepage}
\begin{figure}[ht]
\begin{center}
\epsfig{file=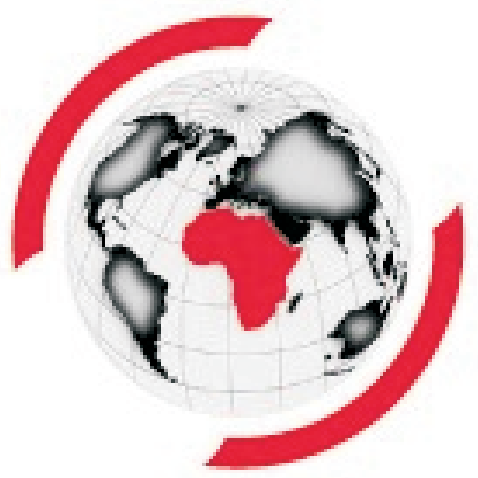,height=4cm}
\end{center}
\end{figure}

\vspace{-3cm}

\begin{flushright}
STIAS-02-002\\
February 2002
\end{flushright}
\begin{centering}

{\ }\vspace{1cm}

{\Large\bf The Cosmological Constant of One-Dimensional}\\

\vspace{7pt}

{\Large\bf Matter Coupled Quantum Gravity is Quantized}\\

\vspace{20pt}

Jan Govaerts\footnote{Permanent address:
Institute of Nuclear Physics, Catholic University
of Louvain, 2, Chemin du Cyclotron, B-1348 Louvain-la-Neuve,
Belgium}$^{,}$\footnote{{\tt govaerts@fynu.ucl.ac.be}}

\vspace{7pt}

{\sl Stellenbosch Institute for Advanced Study (STIAS)\\
Private Bag XI, 7602 Stellenbosch, Republic of South Africa}

\vspace{15pt}

\begin{abstract}

Coupling any interacting quantum mechanical system to gravity in one 
dimension requires the cosmological constant to belong
to the matter energy spectrum and thus to be quantized, even though
the gravity sector is free of any quantum dynamics, while physical states
are also confined to the subspace of matter quantum states whose
energy coincides with the cosmological constant value. These general facts
are illustrated through some simple examples. The physical projector
quantization approach readily leads to the correct representation of
such systems, whereas other approaches relying on gauge fixing methods are
often plagued by Gribov problems in which case the quantization rule
is not properly recovered. Whether such a quantization of the cosmological
constant as well as the other ensuing consequences in terms of physical
states extend to higher dimensional matter-gravity coupled quantum
systems is clearly a fascinating open issue.

\end{abstract}

\vspace{30pt}

{\it Dedicated to John R. Klauder\\
on the occasion of his 70$^{\rm th}$ birthday}

\end{centering}

\end{titlepage}

\setcounter{footnote}{0}

\section{Introduction}
\label{Sect1}

A basic understanding of the gravitational interaction
both in the quantum realm as well as in its relation to the other
fundamental interactions and matter, as witnessed for instance amongst
other equally relevant open issues by the cosmological constant
problem\cite{Weinberg}, is of central importance in all present
attempts at a fundamental unification. Being the example {\sl par
excellence\/} of a system possessing a local gauge symmetry, popular
methods for the quantization of such constrained systems have been
brought to bear on the quantization problem, beginning with pure
gravity. On the other hand, given the insight provided by topological
quantum field theories of gravity\cite{Witten,TQFT}, one is also tempted
to flirt with the idea that geometry emerges in fact only through the
coupling of gravity to the other fundamental interactions and their matter
fields, and that the problems of quantum gravity are to be properly
addressed provided only one also includes these other degrees of freedom.
Whatever the merits of such a proposal, it certainly does not run counter
to all indications motivated by M-theory in this respect.

Given this perspective in the back of one's mind, the aims of the present
note are far more modest by addressing these issues for one-dimensional
gravity only, leading nonetheless to results that hint possibly
at such a connection. Even though gravity is totally free of any classical
or quantum degree of freedom in one dimension, we shall find that when
coupled to whatever quantum matter system, the cosmological constant
must take a quantized value which belongs to the matter energy spectrum, and
that the physical states are restricted to belong to the subspace of
matter quantum states whose energy coincides with the cosmological constant
value. Furthermore, in order to assess the merits of different approaches
to the quantization of gauge invariant systems when applied to theories
of gravity, three such methods are considered explicitly.
It will be shown that Faddeev's reduced phase space
approach\cite{Faddeev,Govx} is unable to properly represent the genuine
quantum dynamics of reparametrization invariant systems, due to Gribov
problems\cite{Gribov,Govx,Gov1,Gov2}. Likewise, the general BRST quantization
methods\cite{BFV,Govx} are to be trusted only for specific gauge fixing
choices which are free of any Gribov problem, while otherwise the correct
quantum dynamics is not recovered either\cite{Gov1,Gov2}. In
contradistinction, the physical projector approach of much more recent
conception\cite{Proj,Klauder}, with in particular its avoidance of any gauge
fixing hence also of any Gribov problem\cite{Gov3}, is shown to be perfectly
capable to correctly represent the actual dynamics of one-dimensional matter
coupled quantum gravity, including the quantized value of the cosmological
constant. Obviously, such conclusions raise the all too intriguing issue of
the extension of these results to matter coupled theories of quantum gravity
in higher spacetime dimensions. In this respect, the physical projector
approach should also prove to be a tool of great efficacy\cite{QG1}.

As a matter of fact, this note grew out from a simple example\cite{Klauder}
of a Gribov problem arising within Faddeev's reduced phase space formulation,
to which the physical projector is simply oblivious. This simple
example turns out to belong to the general class of systems in which
any interacting quantum mechanical matter system is coupled to
one-dimensional gravity. Even though the conclusion that in all such models
the cosmological constant is to be quantized, is totally general and applies
whatever the nature of the interacting quantum matter sector, in the
present note we shall only consider a matter sector described by cartesian
degrees of freedom. Nonetheless, as the reader will readily realize,
the analysis to be presented easily extends to more general cases, such
as for instance quantum mechanical nonlinear sigma models whose degrees
of freedom take values in some curved and/or compact manifold.

The sole mention in the literature known to this author of a 
quantized cosmological constant in the manner discussed here\footnote{The
possibility of a quantized cosmological constant is also mentioned 
for instance in the papers given in Ref.\cite{Other} based on different
considerations.} is that of Ref.\cite{Fujikawa}, in which the interacting
quantum mechanical matter sector is the nonrelativistic hydrogen atom,
again a specific example of the general construction to be discussed
presently.

The note is organized as follows. In Sect.\ref{Sect2}, general
quantum mechanical systems, as yet uncoupled to gravity,
are considered. In Sect.\ref{Sect3}, these systems are coupled
to one-dimensional gravity and then quantized following Dirac's general
approach\cite{Dirac,Govx}, in order to identify their physical sector through 
the physical projector\cite{Proj,Klauder}. It is here that the requirement
of a  quantized cosmological constant arises explicitly. Sect.\ref{Sect4}
then  addresses the same issues, this time from Faddeev's reduced phase space 
approach, to conclude that this method of fixing the reparametrization 
gauge freedom is ill-fated in these systems. In Sect.\ref{Sect5}, the 
Hamiltonian BSRT methods are also brought to bear on the same systems, 
to conclude once again\cite{Gov1,Gov2} that even though these methods lead, 
by construction, to gauge invariant quantities, it is only for an admissible 
gauge fixing procedure that the correct physical content is properly 
identified. Sect.\ref{Sect6} briefly presents some concluding remarks.

\section{The Uncoupled Quantum Matter System}
\label{Sect2}

Quite generally, let us consider an arbitrary classical system whose
Hamiltonian formulation relies on canonical phase space degrees of
freedom $(q^n(t),p_n(t))$ ($n=1,2,\cdots,N$), with the canonical bracket
structure $\{q^n(t),p_m(t)\}=\delta_{n,m}$, whose time evolution is
generated by some Hamiltonian function $H(q^n,p_n)$. The dynamics of such
a system also follows from the variational principle applied to the
first-order phase space Hamiltonian action
\begin{equation}
S\left[q^n,p_n\right]=\int dt\,
\left[\dot{q}^np_n-H(q^n,p_n)\right]\ .
\end{equation}
As indicated previously, we shall assume that the coordinates $q^n(t)$
define a cartesian parametrization of the configuration space of the
system, with $q^n(t)$ and $p_n(t)$ thus taking all possible real values,
even though the construction and its conclusions to be presented in
Sect.\ref{Sect3} remain valid whatever the choice of matter sector.

In the following, when considering examples, we shall restrict to two
specific cases; on the one hand, the single one-dimensional harmonic
oscillator of mass $m$ and angular frequency $\omega$, and on the other hand,
the $N$-dimensional euclidean spherical harmonic oscillator with the
same parameters.

Within this general framework, the classical equations of motion are
\begin{equation}
\dot{q}^n=\frac{\partial H}{\partial p_n}\ \ \ ,\ \ \ 
\dot{p}_n=-\frac{\partial H}{\partial q^n}\ .
\end{equation}
For instance, given initial values $q^n_i=q^n(t_i)$ and $p_{n,i}=p_n(t_i)$,
the solution to these equations is of the form
\begin{equation}
q^n(t)=Q^n(t;q^n_i,p_{n,i})\ \ \ ,\ \ \ 
p_n(t)=P_n(t;q^n_i,p_{n,i})\ ,
\end{equation}
while the ``energy"\footnote{Indeed, this quantity measures the actual
energy of the system only when the time evolution parameter $t$ coincides
with the physical time.} of the system is given by the conserved value
of the Hamiltonian,
\begin{equation}
E(q^n_i,p_{n,i})=H(q^n_i,p_{n,i})\ .
\end{equation}
Here, $Q^n(t;q^n_i,p_{n,i})$ and $P_n(t;q^n_i,p_{n,i})$ are specific
functions of time, also dependent on the initial values for the phase space
degrees of freedom.

As is well known, the Hamiltonian equation of motion for $\dot{q}^n$ may
be used to reduce the conjugate momenta $p_n=p_n(q^n,\dot{q}^n)$ 
and obtain the Lagrangian variational principle for the same dynamics,
\begin{equation}
S[q^n]=\int dt\,L(q^n,\dot{q}^n)\ \ \ ,\ \ \ 
L\left(q^n,\dot{q}^n\right)=\dot{q}^np_n(q^n,\dot{q}^n)-
H\left(q^n,p_n(q^n,\dot{q}^n)\right)\ ,
\end{equation}
as well as the Euler-Lagrange equations of motion
\begin{equation}
\frac{d}{dt}\frac{\partial L}{\partial\dot{q}^n}-
\frac{\partial L}{\partial q^n}=0\ .
\end{equation}
This reduction is especially simple when the conjugate momenta dynamics
separates through an Hamiltonian of the form $H=\sum_{n}p^2_n/(2m_n)+V(q^n)$.
Examples considered further on belong to this latter characterization.

Within this Lagrangian formulation, a choice of boundary values which is
often more convenient than the one above is such that the configuration space
values $q^n(t)$ are specified both at some initial time $t=t_i$ as well as
at some final time $t=t_f$, $q^n_{i,f}=q^n(t_{i,f})$. Nonetheless,
knowledge of the solutions $Q^n(t;q^n_i,p_{n,i})$ and $P_n(t;q^n_i,p_{n,i})$
enables one in principle to construct the solutions to the Euler-Lagrange
equations of motion for this alternative choice of boundary conditions.

The path towards quantization to be taken in this note is that of the
abstract canonical quantization of the above Hamiltonian formulation.
Hence, the quantum system is represented through a linear representation
space ${|\psi>}$ of the abstract Heisenberg algebra
$[\hat{q}^n(t_0),\hat{p}_m(t_0)]=i\hbar\delta_{n,m}$,
equipped with an hermitean inner product $<\,.\,|\,.\,>$ such that
all these operators $\hat{q}^n(t_0)$ and $\hat{p}_n(t_0)$ be---in the best
of cases---self-adjoint\footnote{Canonical quantization is performed in the
Schr\"odinger picture, so that $t_0$ here stands for a reference time at
which quantization is performed for the operator degrees of freedom.}.
Finally, time evolution is generated by Schr\"odinger's equation
\begin{equation}
i\hbar\frac{d}{dt}|\psi,t>=\hat{H}|\psi,t>\,
\end{equation}
where $\hat{H}=\hat{H}(\hat{q}^n(t_0),\hat{p}_n(t_0))$ is a quantum
Hamiltonian operator in correspondence with the classical one, $H(q^n,p_n)$,
and defined in such a manner that it be self-adjoint as well.

For physical consistency, we shall assume that the $\hat{H}$-spectrum
is bounded below. Otherwise, just for the sake of definiteness in our
discussion, we shall further assume that this spectrum be 
discrete\footnote{The quantization of the cosmological constant applies
whether the matter energy spectrum is discrete or continuous, or a mixture
of both.},
\begin{equation}
\hat{H}|E_n,\alpha_n>=E_n|E_n,\alpha_n>\ \ ,\ \ \alpha_n=1,2,\cdots,d_n\ \ ,
\ \ n=0,1,2,\cdots\ ,
\end{equation}
with a degeneracy $d_n\ge 1$ at each of the energy levels $E_n$ 
($n=0,1,2,\cdots)$ labelled by $\alpha_n=1,2,\cdots,d_n$,
and with a choice of orthonormalized energy eigenstates
\begin{equation}
<E_n,\alpha_n|E_m,\alpha_m>=\delta_{n,m}\,\delta_{\alpha_n,\alpha_m}\ ,
\end{equation}
which implies
\begin{equation}
\one=\sum_{n,\alpha_n}|E_n,\alpha_n><E_n,\alpha_n|\ .
\end{equation}
Consequently, the Hamiltonian operator possesses the spectral representation
\begin{equation}
\hat{H}=\sum_{n,\alpha_n}|E_n,\alpha_n>E_n<E_n,\alpha_n|\ ,
\end{equation}
while the time evolution operator 
$\hat{U}(t_2,t_1)=e^{-i/\hbar(t_2-t_1)\hat{H}}$ of the quantum system
simply reads
\begin{equation}
\hat{U}(t_2,t_1)=\sum_{n,\alpha_n}|E_n,\alpha_n>\,
e^{-i/\hbar(t_2-t_1)E_n}\,<E_n,\alpha_n|\ .
\end{equation}
Hence, given any initial state
\begin{equation}
|\psi,t=t_i>=\sum_{n,\alpha_n}|E_n,\alpha_n>\psi_{n,\alpha_n}\ \ \ ,\ \ \ 
\psi_{n,\alpha_n}=<E_n,\alpha_n|\psi,t=t_i>\ ,
\end{equation}
the general solution to Schr\"odinger's equation is
\begin{equation}
|\psi,t>=\hat{U}(t,t_i)|\psi,t_i>=
\sum_{n,\alpha_n}|E_n,\alpha_n>e^{-i/\hbar E_n(t-t_i)}\psi_{n,\alpha_n}\ .
\end{equation}

It should be clear how any (of the well known) quantum mechanical system(s)
may be brought into correspondence with such a general description, beginning
with the ordinary harmonic oscillator in its Fock space representation.

\section{The Gravity Coupled System}
\label{Sect3}

Given any general mechanical system as described in Sect.\ref{Sect2}, let
us now consider the following first-order action principle,
\begin{equation}
S[q^n,p_n;\lambda]=\int\,dt\,
\left[\dot{q}^np_n-\lambda\left(H(q^n,p_n)-\Lambda\right)\right]\ ,
\label{eq:Scoupled}
\end{equation}
where $\lambda(t)$ is an arbitrary function of $t$ and $\Lambda$ an
arbitrary real constant parameter. Clearly, $\lambda(t)$ is a Lagrange
multiplier for a constraint on phase space, namely $\phi(q^n,p_n)=0$, with
\begin{equation}
\phi(q^n,p_n)=H(q^n,p_n)-\Lambda\ ,
\end{equation}
while the Hamiltonian equations of motion now read
\begin{equation}
\dot{q}^n=\lambda\frac{\partial H}{\partial p_n}\ \ \ ,\ \ \ 
\dot{p}_n=-\lambda\frac{\partial H}{\partial q^n}\ ,
\label{eq:mt}
\end{equation}
whose solutions are subjected to the constraint $H(q^n,p_n)=\Lambda$.
As a matter of fact, (\ref{eq:Scoupled}) defines a constrained
system\cite{Dirac,Govx} whose phase space coordinates $(q^n,p_n)$ are 
canonical, $\{q^n,p_m\}=\delta_{n,m}$, whose time evolution is generated 
by the total Hamiltonian $H_T=\lambda(H-\Lambda)$, and which possesses the
gauge invariance degree of freedom under arbitrary local reparametrizations
in $t$ generated by the first-class constraint $\phi=H-\Lambda$. The
associated first-class Hamiltonian vanishes identically, as befits
any reparametrization invariant system, so that the total Hamiltonian which
generates time dependency is indeed ``pure gauge", $H_T=\lambda\phi$,
while the constraint $\phi=0$ that classical solutions must meet expresses 
their gauge invariance under local time reparametrizations. The fact that 
(\ref{eq:Scoupled}) does define the coupling of the matter system
described by $H(q^n,p_n)$ to gravity in a one-dimensional ``spacetime"
with $\Lambda$ being a cosmological constant may be justified from 
complementary viewpoints. In particular, $\lambda(t)$ is nothing but a
world-line einbein, with $\lambda^2(t)$ defining an intrinsic world-line
metric\footnote{Note that the classes of models considered here include
the scalar relativistic particle propagating in a spacetime of whatever
geometry, the parameter $\Lambda$ being directly related to the particle's
squared mass which must indeed belong to the ``energy" spectrum of the
operator $\hat{P}_\mu\hat{P}^\mu$.}.

First, given initial values $q^n_i$ and $p_{n,i}$ at $t=t_i$
for the phase space degrees of freedom, let us construct the general
solutions to the equations of motion (\ref{eq:mt}) whatever the choice for 
the Lagrange multiplier function $\lambda(t)$. For this purpose, consider
the function $\tau(t)$, with $\tau_i=t_i$, such that
\begin{equation}
\tau-\tau_i=\int_{t_i}^t\,dt'\,\lambda(t')\ \ \ ,\ \ \ 
\tau(t_i)=t_i=\tau_i\ .
\end{equation}
It then follows that in terms of this variable $\tau$,
the equations of motion (\ref{eq:mt}) reduce to those of the uncoupled
matter system,
\begin{equation}
\frac{d q^n}{d\tau}=\frac{\partial H}{\partial p_n}\ \ \ ,\ \ \ 
\frac{d p_n}{d\tau}=-\frac{\partial H}{\partial q^n}\ ,
\end{equation}
while the first-order action (\ref{eq:Scoupled}) reads
\begin{equation}
S[q^n,p_n;\lambda]=\int\,d\tau\,
\left[\frac{dq^n}{d\tau}p_n-(H(q^n,p_n)-\Lambda)\right]\ .
\end{equation}
Consequently, irrespective of the choice for $\lambda(t)$, the general
solution to the equations of motion associated to the initial values 
$(q^n_i,p_{n,i})$ is of the same form as that for the uncoupled system,
\begin{equation}
q^n(t)=Q^n(\tau(t);q^n_i,p_{n,i})\ \ \ ,\ \ \ 
p^n(t)=P^n(\tau(t);q^n_i,p_{n,i})\ ,
\end{equation}
with however the time dependency on $t$ substituted by that on the parameter
$\tau(t)$, and the further restriction that the ``energy" $E$ of the solution
must coincide with the cosmological constant value $\Lambda$,
\begin{equation}
E(q^n_i,p_{n,i})=H(q^n_i,p_{n,i})=\Lambda\ .
\end{equation}
In other words, irrespective of the choice of world-line (re)parametrization 
characterized by the function $\tau(t)$ such that
$\tau(t_i)=t_i$, and thus associated to the Lagrange multiplier
$\lambda(t)=d\tau(t)/dt$, the solution is given by the same functions
$Q^n(\tau;q^n_i,p_{n,i})$ and $P_n(\tau;q^n_i,p_{n,i})$ as those of
the uncoupled system, with however the restriction that the cosmological
constant $\Lambda$ must belong to the $H$-spectrum and that the energy $E$
of the solution must coincide with $\Lambda$. The dependency of the general
solution to the equations of motion on an arbitrary function $\tau(t)$ of
time is the manifest hallmark of a system which is gauge invariant under local
reparametrizations in one dimension, namely in the time parameter. The
physical content of such systems is thus that which is totally independent
of the world-line parametrization, as are for example the relations which
exist among  the phase space degrees of freedom $q^n$ and $p_n$ independently
of $t$ but dependent on the initial values $q^n_i$ and $p_{n,i}$, such as for
instance the energy $E(q^n_i,p_{n,i})=\Lambda$.

The fact that we are indeed dealing with a theory of one-dimensional gravity
coupled to matter dynamics may also be justified from another viewpoint.
Applying the Hamiltonian reduction of the conjugate momenta $p_n$ discussed
in Sect.\ref{Sect2} but based this time on the coupled action 
(\ref{eq:Scoupled}) and the associated equations of motion (\ref{eq:mt}),
one readily concludes that the Lagrangian formulation of the coupled
system is,
\begin{equation}
S[q^n;\lambda]=\int d\tau\,\left[L\left(q^n,\frac{dq^n}{d\tau}\right)+
\Lambda\right]=\int dt\,\lambda(t)\,\left[L\left(q^n,\frac{1}{\lambda(t)}
\frac{dq^n}{dt}\right)+\Lambda\right]\ .
\end{equation}
In this form, it is clear that we are indeed dealing with an intrinsic
world-line metric defined by the invariant line element
\begin{equation}
ds^2=dt^2\lambda^2(t)=d\tau^2\ ,
\end{equation}
with $\lambda(t)$ and $\lambda^2(t)$ as einbein and metric structures, 
coupled in a reparametrization invariant way to the matter 
Lagrangian as well as to a world-line cosmological term $\Lambda$.

Finally, let us consider the local gauge invariance properties in the
Hamiltonian formulation. Since $\phi=H-\Lambda$ is the generator for
the associated infinitesimal transformations in phase space, their explicit
expression reads
\begin{equation}
\delta_\epsilon q^n=\epsilon\,\left\{q^n,\phi\right\}=\epsilon\,
\frac{\partial H}{\partial p_n}\ \ ,\ \ 
\delta_\epsilon p_n=\epsilon\,\left\{p^n,\phi\right\}=-\epsilon\,
\frac{\partial H}{\partial q^n}\ \ ,\ \ 
\delta_\epsilon\lambda=\dot{\epsilon}\ ,
\label{eq:transf}
\end{equation}
where $\epsilon(t)$ is some arbitrary function. Indeed, the variation
of the first-order action (\ref{eq:Scoupled}) is then a surface term,
\begin{equation}
\delta_\epsilon S[q^n,p_n;\lambda]=\int dt\,\frac{d}{dt}
\left[\,\epsilon\left(p_n\frac{\partial H}{\partial p_n}-H+\Lambda\right)
\right]\ ,
\end{equation}
thus expressing the fact that the transformations (\ref{eq:transf}) do
define a symmetry of the system, namely world-line reparametrization
invariance of a general matter coupled one-dimensional gravity system.

The abstract canonical quantization of the system is straightforward
enough. The gravitational sector being dynamics free, the space of quantum 
states is that of the matter sector, as described in Sect.\ref{Sect2},
whereas no further quantum operators are associated to the gravitational
sector. In particular, the arbitrary Lagrange multiplier function
$\lambda(t)$ is not related to a quantum operator, and still serves the 
sole purpose of parametrizing the gauge freedom related to the choice of 
world-line parametrization. Compared to the uncoupled quantum matter system,
the only modification is that time dependency of states is generated through 
a Schr\"odinger equation which now reads
\begin{equation}
i\hbar\frac{d}{dt}|\psi,t>=\lambda(t)\left[\hat{H}-\Lambda\right]\,|\psi,t>\ ,
\end{equation}
or equivalently, in terms of the function $\tau(t)$,
\begin{equation}
i\hbar\frac{d}{d\tau}|\psi,t(\tau)>=\left[\hat{H}-\Lambda\right]\,
|\psi,t(\tau)>\ ,
\end{equation}
thus making manifest yet again the fact that $\lambda(t)$ indeed parametrizes
the freedom in world-line parametrizations.

Given an arbitrary initial state 
$|\psi,t_i>=\sum_{n,\alpha_n}|E_n,\alpha_n>\psi_{n,\alpha_n}$, the solution
to the Schr\"odinger equation is thus now
\begin{equation}
|\psi,t>=\sum_{n,\alpha_n}|E_n,\alpha_n>\,
e^{-i/\hbar(E_n-\Lambda)\int_{t_i}^tdt'\lambda(t')}\psi_{n,\alpha_n}\ .
\end{equation}
More generally, the quantum time evolution operator of the quantized
system is
\begin{equation}
\hat{U}(t_2,t_1)=\sum_{n,\alpha_n}|E_n,\alpha_n>\,e^{-i/\hbar(E_n-\Lambda)\
\int_{t_1}^{t_2}dt\lambda(t)}\,<E_n,\alpha_n|\ .
\end{equation}

However, all quantum states of the system may not be regarded as being
physical, {\sl i.e.\/} gauge invariants states, but only those\cite{Dirac} 
that are annihilated by the gauge generator $\hat{\phi}=\hat{H}-\Lambda$,
\begin{equation}
\left[\hat{H}-\Lambda\right]\,|\psi_{\rm phys}>=0\ .
\end{equation}
When combined with the Schr\"odinger equation, note that this restriction
implies that physical states are independent of time, namely they
are indeed independent of the world-line parametrization as it should.

On the other hand, we also recover for quantum physical states a situation
identical to that for the classical gauge invariant solutions, namely the fact
that in order for the physical content of the system not to be void,
the cosmological constant $\Lambda$ must belong to the $\hat{H}$-spectrum
of the matter sector, namely
\begin{equation}
\Lambda=E_{n_0}\ ,
\end{equation}
for some specific positive integer value of $n_0$. In other words,
one-dimensional quantum gravity coupled to interacting quantum matter is
physical provided only the cosmological constant is quantized with a value
belonging to the energy spectrum of the quantum matter sector. Clearly, this
result is very general, whatever the quantum matter sector. If the latter
possesses only a discrete spectrum, $\Lambda$ is quantized within that
discrete spectrum, and likewise for a continuous domain in the
$\hat{H}$-spectrum.

Let us hence assume that the value $\Lambda$ coincides with one of the
energy eigenvalues $E_{n_0}$. Consequently, the subspace of physical
quantum states is spanned by the states $|E_{n_0},\alpha_{n_0}>$
with a degeneracy $d_{n_0}$,
\begin{equation}
|\psi_{\rm phys}>=\sum_{\alpha_{n_0}}|E_{n_0},\alpha_{n_0}>\,
\psi_{\alpha_{n_0}}\ \ ,\ \ 
\psi_{\alpha_{n_0}}=<E_{n_0},\alpha_{n_0}|\psi_{\rm phys}>\ ,
\end{equation}
to which the following physical projector\cite{Proj,Klauder} is thus
associated,
\begin{equation}
\proj=\sum_{\alpha_{n_0}}
|E_{n_0},\alpha_{n_0}><E_{n_0},\alpha_{n_0}|\ \ ,\ \ 
\proj\,^2=\proj\ \ ,\ \ \proj\,^\dagger=\proj\ .
\end{equation}
In particular, the physical time evolution operator on the physical
subspace simply reduces to
\begin{equation}
\hat{U}_{\rm phys}(t_2,t_1)=\proj\,\hat{U}(t_2,t_1)\,\proj=
\sum_{\alpha_{n_0}}
|E_{n_0},\alpha_{n_0}><E_{n_0},\alpha_{n_0}|=\proj\ .
\end{equation}
Once again, this result expresses the world-line reparametrization gauge
invariance fact that quantum physical states are time independent, as befits
any physical state of a quantum theory for gravity. Furthermore, it
illustrates the general feature that for reparametrization invariant quantum 
theories, the physical projector embodies all the physical content of matter 
coupled quantum gravity. In one dimension, the physics of any such system lies
within the subspace of quantum matter states whose energy coincides with
the cosmological constant. It certainly is a fascinating issue to
determine how this conclusion extends to higher dimensional matter
coupled quantum gravity theories.

The above characterization of the quantum physical subspace is also
reminiscent of topological quantum field theories\cite{Witten,TQFT}, 
namely theories whose reparametrization invariant physical content is solely
dependent on the topology of the underlying manifold, with in particular
a finite dimensional space of quantum states. One could debate whether
the matter coupled quantum gravity systems of this section define the
simplest examples of topological quantum field theories, since the
considered world-line topology remains trivial, but they certainly
provide the simplest examples of quantum gravity systems with a space
of quantum physical states which is finite dimensional and such that the
cosmological constant is necessarily quantized in a manner dependent
on the quantum matter dynamics to which gravity is coupled, even though
the gravitational sector is totally trivial.

\section{Faddeev's Reduced Phase Space Formulation}
\label{Sect4}

Given that the above quantization of one-dimensional matter coupled quantum
gravity is straightforward and free of any ambiguity, including the complete
characterization of its physical sector and of its quantum dynamics
through the physical projector,
it is interesting to confront these results with those that follow from 
alternative approaches to the quantization of constrained dynamics, which all 
rely of some method to fix the gauge freedom associated to first-class
constraints. Often, such gauge fixing procedures run into Gribov 
problems\cite{Gribov} of one type or another\cite{Gov2}, rendering the
physical interpretation difficult since the quantized system is then
no longer physically equivalent to the one originally considered. 
In contradistinction, the previous approach solely based on the physical 
projector but not on any gauge fixing procedure whatsoever, is guaranteed 
to be free of any Gribov ambiguity\cite{Gov3}, and thus to truly represent
the actual quantum dynamics of the original system.

This section considers the application to the previous systems of a quite
popular gauge fixing procedure, namely Faddeev's reduced phase space
approach\cite{Faddeev,Govx}. This shall be done explicitly for the spherical
harmonic oscillator in $N$ dimensions, defined by
\begin{equation}
H(\vec{q},\vec{p}\,)=\frac{1}{2m}\vec{p}\,^2+\frac{1}{2}m\omega^2\vec{q}\,^2\ ,
\end{equation}
but once again, the conclusions will be seen to remain valid in
general. Two distinct gauge fixing conditions will
be considered, both leading to the conclusion that for this general
class of systems, Faddeev's approach is unable to provide their physically
correct quantization as described in Sect.\ref{Sect3}.

The reason why this specific choice of matter sector is made, is that
the infinitesimal Hamiltonian world-line reparametrizations (\ref{eq:transf})
may then readily be extended to finite transformations, given by
\begin{equation}
\begin{array}{r c l}
\vec{q}\,'(t)&=&\vec{q}(t)\,\cos\omega\epsilon(t)+
\frac{\vec{p}(t)}{m\omega}\,\sin\omega\epsilon(t)\ ,\\
\vec{p}\,'(t)&=&-m\omega\vec{q}(t)\,\sin\omega\epsilon(t)+
\vec{p}(t)\,\cos\omega\epsilon(t)\ ,\\
\lambda'(t)&=&\lambda(t)+\dot{\epsilon}(t)\ ,
\end{array}
\end{equation}
$\epsilon(t)$ being an arbitrary function, possibly subjected to
boundary conditions depending on the choice of boundary conditions
on the phase space variables $\vec{q}(t)$ and $\vec{p}(t)$.
A straightforward substitution into the first-order action
$S[\vec{q},\vec{p};\lambda]=\int dt[\dot{\vec{q}}\cdot\vec{p}-
\lambda(H-\Lambda)]$ then finds
\begin{equation}
S[\vec{q}\,',\vec{p}\,';\lambda']=S[\vec{q},\vec{p};\lambda]+
\int dt\,\frac{d}{dt}
\left[-\vec{q}\cdot\vec{p}\sin^2\omega\epsilon+
\left(\frac{\vec{p}\,^2}{2m\omega}-\frac{1}{2}m\omega\vec{q}\,^2\right)
\sin\omega\epsilon\cos\omega\epsilon+\Lambda\,\epsilon\right]\ ,
\end{equation}
clearly displaying the gauge invariance of the system under local
world-line reparametrizations in phase space. Note that for the
present system, these transformations coincide with rotations between
the configuration space variables $\vec{q}$ and their conjugate momenta
$\vec{p}$, a property related to the fact that energy conservation restricts 
the dynamics onto an invariant $N$-dimensional torus in phase space.

The latter remark also shows that when this matter system is coupled to
one-dimensional gravity, through\footnote{In this discussion, we take
$\Lambda$ to be strictly positive, $\Lambda>0$. A vanishing cosmological 
constant, $\Lambda=0$, is possible only for the trivial configuration 
$\vec{q}(t)=\vec{0}$, $\vec{p}(t)=\vec{0}$, while no solutions exist for
a negative value of $\Lambda$.}
\begin{equation}
H_T=\lambda(t)\left[\frac{1}{2m}\vec{p}\,^2+\frac{1}{2}m\omega^2\vec{q}\,^2-
\Lambda\right]\ \ \ ,\ \ \ \Lambda>0\ ,
\end{equation}
whatever the configuration solving the equations of motion,
there always exists a finite gauge transformation $\epsilon(t)$ such
that one of the position (or momentum) degrees of freedom vanishes
at all times, say $q^1(t)=0$. This property thus suggests to consider
a Faddeev reduced phase space formulation of the system defined for
instance by the gauge fixing condition\cite{Klauder}
\begin{equation}
\Omega=q^1=0\ .
\end{equation}
One of the purposes of such a gauge fixing condition is to single out
a specific Lagrange multiplier function, namely
a specific world-line parametrization, by requiring this condition to
be stable under time evolution. Given the present choice as well as
arbitrary values for\footnote{The solution $\lambda(t)=0$ may be avoided only
if $p_1(t)=0$ at all times, which in general is inconsistent with an
arbitrary choice of initial values at $t=t_i$. It is only if 
both $q^1_i=0$ as well as $p_{1,i}=0$ that the $n=1$ degree of freedom
decouples entirely from the dynamics.}
$p_1(t)\ne 0$, this implies $\lambda(t)=0$, namely a singular world-line
parametrization such that $\tau(t)=t_i$ at all times!
Proceeding nonetheless, one finds that the reduced system is represented by
the degrees of freedom $(q^i,p_i)$ with $i=2,3,\cdots,N$ whose Dirac
brackets are still given by the canonical brackets
$\{q^i(t),p_j(t)\}=\delta_{ij}$, while
the constraint $\phi=H-\Lambda=0$ and the gauge fixing condition
$\Omega=0$ are solved by
\begin{equation}
q^1=0\ \ \ ,\ \ \ 
p_1=\pm\sqrt{2m\left[\Lambda-\sum_i\left(\frac{p^2_i}{2m}+
\frac{1}{2}m\omega^2{q^i}^2\right)\right]}\ .
\end{equation}
Furthermore, given the value $\lambda(t)=0$, the effective Hamiltonian
which generates time evolution on reduced phase space vanishes identically,
$H_{\rm red}=0$.

Consequently, upon canonical quantization of this reduced phase space
formulation of the system, it is clear that no quantization condition
whatsoever on the cosmological constant $\Lambda$ may possibly arise,
while at the same time the quantum space of states cannot coincide
with the finite dimensional subspace of quantum states of the $N$-dimensional 
harmonic oscillator whose energy coincides with a quantized 
value for the cosmological constant $\Lambda=E_{n_0}=\hbar\omega(n_0+N/2)$,
namely specifically\cite{Klauder} the totally symmetric $SU(N)$
representation of dimension $d_{n_0}=(N+n_0-1)!/(n_0!(N-1)!)$, as established
in general terms  in Sect.\ref{Sect3}. In other words, Faddeev's reduced
phase space quantization of the system  associated to the gauge fixing
condition $q^1=0$ leads to a system whose physics is totally different from
that of the original reparametrization invariant system\footnote{In fact,
following arguments similar to those developed in Ref.\cite{Gov2} but which
shall not be pursued here, the reason for this lack of equivalence may be
traced back\cite{Klauder} to Gribov problems of the  first and second types
associated to the gauge fixing condition $q^1=0$.}, a fact which applies in
general whenever a given gauge fixing procedure suffers Gribov
problems\cite{Gov2}.

Thinking that this conclusion could possibly be related to the singular 
world-line parametrization such that $\lambda(t)=0$ which is
associated to the choice $q^1=0$, one may consider yet another gauge fixing 
condition, based on the classical solutions to the Hamiltonian equations of 
motion. Given initial values $\vec{q}_i=\vec{q}(t_i)$ and 
$\vec{p}_i=\vec{p}(t_i)$, the general solution is
\begin{equation}
\vec{q}(t)=\vec{q}_i\cos\omega(t-t_i)+\frac{1}{m\omega}\vec{p}_i
\sin\omega(t-t_i)\ \ ,\ \ 
\vec{p}(t)=\vec{p}_i\cos\omega(t-t_i)-m\omega\vec{q}_i
\sin\omega(t-t_i)\ \ .
\end{equation}
For the gravity coupled system, let us thus consider the gauge fixing
condition
\begin{equation}
\Omega=q^1-\left[q^1_i\cos\omega(t-t_i)+
\frac{p_{1,i}}{m\omega}\sin\omega(t-t_i)\right]\ .
\end{equation}
Correspondingly, the world-line parametrization is characterized by
the Lagrange multiplier
\begin{equation}
\lambda(t)=\frac{1}{p_1(t)}\left[p_{1,i}\cos\omega(t-t_i)-
m\omega q^1_i\sin\omega(t-t_i)\right]\ ,
\end{equation}
while the reduced phase space variables are the coordinates
$(q^i,p_i)$ with $i=2,3,\cdots,N$ whose Dirac brackets remain
canonical\footnote{The reduced phase space Hamiltonian is not given here.}, 
with the following solutions to the first-class constraint
$\phi=0$ and gauge fixing condition $\Omega=0$,
\begin{equation}
q^1(t)=q^1_i\cos\omega(t-t_i)+\frac{p_{1,i}}{m\omega}\sin\omega(t-t_i)\ \ ,\ \ 
p_1(t)=\pm\sqrt{2m\left[\Lambda-\sum_i\frac{p^2_i}{2m}-
\frac{1}{2}m\omega^2\vec{q}\,^2\right]}\ .
\end{equation}
Clearly, even though this gauge fixing is nonsingular given the world-line 
parametrization with $\tau(t)=t_i+\int_{t_i}^t dt'\lambda(t')$, canonical 
quantization of this formulation of the system cannot recover either its
correct quantum physical content as established in Sect.\ref{Sect3} with in 
particular the requirement of a quantized cosmological constant $\Lambda$.
Once again\cite{Gov1,Gov2}, Faddeev's reduced phase space approach fails for 
these general classes of one-dimensional reparametrization invariant systems.
We are forced to such a conclusion for both examples of gauge fixing 
conditions above, even though the Faddeev-Popov determinant 
$\{\Omega,\phi\}=p_1/m$ does not vanish in either case. Contrary to what is 
often claimed, though necessary, a nonvanishing Faddeev-Popov determinant is
not a sufficient condition for an admissible gauge fixing free of Gribov
ambiguities.

It would also be possible to consider these difficulties of Faddeev's
reduced phase space approach from the path integral viewpoint, but this
issue shall be left aside in this note.

\section{Hamiltonian BRST Quantization}
\label{Sect5}

Another quite popular approach to constrained dynamics through gauge fixing is 
the BRST invariant one, namely the BFV-BRST Hamiltonian formalism\cite{BFV}
appropriate to the canonical quantization path taken in this note\cite{Govx}.
Here for simplicity, this quantization procedure is applied to the
one-dimensional harmonic oscillator, whose quantization is best represented
in Fock space  through creation and annihilation operators $a$ and $a^\dagger$
such that  $[a,a^\dagger]=\one$ and whose quantum Hamiltonian is
\begin{equation}
\hat{H}=\hbar\omega\left[a^\dagger a+\frac{1}{2}\right]\ .
\end{equation}
Correspondingly, the orthonormalized Fock basis is spanned by the vectors
$|n>=(a^\dagger)^n|0>/\sqrt{n!}$ with $<n|m>=\delta_{nm}$, which are
$\hat{H}$-eigenstates, $\hat{H}|n>=E_n|n>$, with $E_n=\hbar\omega(n+1/2)$, 
$(n=0,1,2,\cdots)$.

In the Hamiltonian BFV-BRST approach for the gravity coupled system, 
phase space is extended by introducing a canonical momentum $p_\lambda$ 
conjugate to the Lagrange multiplier $\lambda$, with bracket 
$\{\lambda,p_\lambda\}=1$, as well as canonical conjugate pairs of ghost 
degrees of freedom $\eta^a$ and ${\cal P}_a$ of Grassmann odd parity 
associated to the first-class constraints $G_a=(p_\lambda,\phi)$ $(a=1,2)$ 
with brackets $\{\eta^a,{\cal P}_b\}=-\delta^a_b$. Local gauge invariance is 
then traded for global BRST invariance generated by the nilpotent BRST charge 
which for the present system reads
\begin{equation}
Q_B=\eta^1p_\lambda+\eta^2\left[\frac{p^2}{2m}+\frac{1}{2}m\omega^2q^2-
\Lambda\right]\ \ \ ,\ \ \
\{Q_B,Q_B\}=0\ .
\end{equation}
The ghosts $\eta^a$ carry ghost number $(+1)$ and are real under
complex conjugation, whereas the ghosts ${\cal P}_a$ carry ghost number $(-1)$
and are pure imaginary under complex conjugation.

Time evolution in this extended phase space is generated by a BRST
invariant extension of the gauge invariant first-class Hamiltonian
of the system. In our case, the general BRST invariant Hamiltonian
is of the form
\begin{equation}
H_{\rm eff}=-\{\Psi,Q_B\}\ ,
\end{equation}
where $\Psi$ is some {\sl a priori\/} arbitrary function defined over 
extended phase space, which is pure imaginary under complex conjugation, is 
of ghost number $(-1)$, and is of odd Grassmann parity. As a matter of fact, 
gauge fixing of the system is performed through some choice for this function 
$\Psi$.

Taking for instance
\begin{equation}
\Psi=F(\lambda){\cal P}_1+\lambda{\cal P}_2\ ,
\label{eq:Psi}
\end{equation}
where $F(\lambda)$ is some function of the Lagrange multiplier,
leads to the BRST invariant Hamiltonian
\begin{equation}
H_{\rm eff}=\lambda\left[\frac{p^2}{2m}+\frac{1}{2}m\omega^2q^2-\Lambda\right]+
F(\lambda)p_\lambda-F'(\lambda){\cal P}_1\eta^1-{\cal P}_2\eta^1\ ,
\end{equation}
and in turn to BRST invariant equations of motion which include
\begin{equation}
\dot{\lambda}=F(\lambda)\ .
\end{equation}
Hence, a given choice for the function $F(\lambda)$ and for the initial
(or final) value $\lambda_i=\lambda(t_i)$ leads to a specific
solution $\lambda(t)$, namely a specific world-line parametrization or
gauge fixing of the system.

Choices free of Gribov ambiguities are\cite{Gov1,Gov2} 
$F(\lambda)=\alpha\lambda+\beta$ where $\alpha$, $\beta$ are arbitrary 
constant parameters. On the other hand, examples of gauge fixing
conditions leading to Gribov problems are\cite{Gov1,Gov2}
$F(\lambda)=\alpha\lambda^2+\beta\lambda+\delta$, $F(\lambda)=\alpha\lambda^3$
or $F(\lambda)=e^{-\alpha\lambda}$. Even though, by construction, the 
formulation of the system is BRST and thus gauge invariant whatever the
choice for $\Psi$ or $F(\lambda)$, the physics that is being described,
albeit gauge invariant, does depend on the gauge fixing condition. It is only
when the latter is free of any Gribov problem that an admissible gauge
fixing has been achieved, and that a description of the system's dynamics
which is equivalent to the original one is recovered. The same statement
applies already at the classical level\cite{Gov2} for the solutions to the
BRST invariant equations of motion given a choice of BRST invariant boundary
conditions. It is only for an admissible choice of gauge fixing function
$\Psi$ that the correct classical solutions are recovered, and only those.

By considering the BRST transformations of the extended phase space
degrees of freedom, one realizes that a general choice of BRST
invariant boundary conditions, irrespective of the boundary
values for the original configuration space coordinates $q^n$
at $t=t_i$ and $t=t_f$, is such that
\begin{equation}
p_{\lambda}(t_{i,f})=0\ \ ,\ \ 
{\cal P}_1(t_{i,f})=0\ \ ,\ \ 
\eta^2(t_{i,f})=0\ .
\label{eq:BRSTbc}
\end{equation}
It is thus also for such a choice of BRST invariant external quantum
states that the BRST invariant quantum evolution operator is to be
considered, namely that these matrix elements are to represent the gauge 
invariant time evolution of the quantum system given initial and final
$\hat{q}^n$-eigenstate configurations $q^n_i$ and $q^n_f$, to which
physical states only contribute both as intermediate as well as external
states.

Canonical quantization of the BFV-BRST extended system is 
straightforward\cite{Govx}. In the Grassmann even sector, one has the
familiar operator algebras $[a,a^\dagger]=\one$ and 
$[\hat{\lambda},\hat{p}_\lambda]=i\hbar$. In the ghost sector, one has
the fermionic algebra $\{\hat{c}^a,\hat{b}_b\}=\delta^a_b$,
with $\hat{c}^a=\hat{\eta}^a$ and $\hat{b}_a=i\hat{\cal P}_a/\hbar$,
and the properties $\hat{c}^{a\dagger}=\hat{c}^a$ and
${\hat{b}_a}^\dagger=\hat{b}_a$. The representation of this $(b,c)$
algebra is spanned by a four-dimensional Hilbert space with
basis vectors $|\pm\pm>$ such that
\begin{equation}
\begin{array}{l l l l}
\hat{c}^1|-->=|+->\ ,\
&\hat{c}^1|+->=0\ ,\
&\hat{c}^1|-+>=|++>\ ,\
&\hat{c}^1|++>=0\ ,\\
\hat{c}^2|-->=|-+>\ ,\
&\hat{c}^2|+->=-|++>\ ,\
&\hat{c}^2|-+>=0\ ,\
&\hat{c}^2|++>=0\ ,\\
\hat{b}_1|-->=0\ ,\
&\hat{b}_1|+->=|-->\ ,\
&\hat{b}_1|-+>=0\ ,\
&\hat{b}_1|++>=|-+>\ ,\\
\hat{b}_2|-->=0\ ,\
&\hat{b}_2|+->=0\ ,\
&\hat{b}_2|-+>=|-->\ ,\
&\hat{b}_2|++>=-|+->\ ,
\end{array}
\end{equation}
and whose inner product is defined through the following only nonvanishing
matrix elements,
\begin{equation}
<--|++>=-\alpha\ \ ,\ \ 
<+-|-+>=-\alpha\ \ ,\ \ 
<-+|+->=\alpha\ \ ,\ \ 
<++|-->=\alpha\ ,
\end{equation}
with $\alpha=\pm i$. Finally, the antihermitean ghost number operator reads
\begin{equation}
\hat{Q}_c=\frac{1}{2}\left[\hat{c}^a\hat{b}_a-\hat{b}_a\hat{c}^a\right]\ ,
\end{equation}
so that $|-->$ is of minimal ghost number $(-1)$, $|+->$ and
$|-+>$ are both of vanishing ghost number, and $|++>$ is of
maximal ghost number $(+1)$. 

As is characteristic of the BRST construction, physical gauge invariant states
correspond to BRST invariant states of minimal ghost number, thus\footnote{As
we shall see, upon imposing Schr\"odinger's equation,
physical states are also time independent.}
\begin{equation}
\hat{Q}_B|\psi_{\rm phys},t>=0\ \ \ ,\ \ \ 
\hat{Q}_c|\psi_{\rm phys},t>=-|\psi_{\rm phys},t>\ ,
\end{equation}
where the nilpotent quantum BRST operator is
\begin{equation}
\hat{Q}_B=\hat{c}^1\hat{p}_\lambda+\hat{c}^2
\left[\hbar\omega(a^\dagger a+\frac{1}{2})-\Lambda\right]\ \ \ ,\ \ \ 
\hat{Q}^2_B=0\ .
\end{equation}
Consequently once again, in order to obtain a quantum theory with physical 
content, it is necessary that the cosmological constant $\Lambda$
belongs to the matter energy spectrum, namely in the present case
\begin{equation}
\Lambda=\hbar\omega(n_0+\frac{1}{2})\ ,
\end{equation}
for some positive integer value $n_0$, a condition which we shall henceforth
assume to be met. The general solution for physical states
is then of the form
\begin{equation}
|\psi_{\rm phys},t>=|n_0>\otimes|p_\lambda=0>\otimes|-->\,\psi_{--,n_0}(t)\ ,
\label{eq:physicalBRST}
\end{equation}
$\psi_{--,n_0}(t)$ being an arbitrary complex function of time,
which turns out to be a constant when solving the Schr\"odinger equation
hereafter. Here, $|p_\lambda>$ denotes the $\hat{p}_\lambda$-eigenstate
basis normalized such that 
$<p_\lambda|p_\lambda'>=\delta(p_\lambda-p_\lambda')$.

As a matter of fact, it is possible to solve for all BRST invariant states
irrespective of their ghost number, $\hat{Q}_B|\psi,t>=0$, leading to the 
following general decomposition,
\begin{equation}
\begin{array}{r c c l}
|\psi,t>&=&&|n_0>\otimes|p_\lambda=0>\otimes|-->\,\psi_{--,n_0}(t) \\
&&+&|n_0>\otimes|p_\lambda=0>\otimes|+->\,\psi_{+-,n_0}(t) \\
&&+&|n_0>\otimes|p_\lambda=0>\otimes|-+>\,\psi_{-+,n_0}(t) \\
&&+&|n_0>\otimes|p_\lambda=0>\otimes|++>\,\psi_{++,n_0}(t)\
+\ \hat{Q}_B|\chi,t>\ ,
\end{array}
\label{eq:genBRST}
\end{equation}
where $|\chi,t>$ is an arbitrary quantum state. In this form, the
BRST cohomology classes are explicit as well, showing that there
are four distinct nontrivial BRST invariant cohomology classes, each
of dimension unity. Besides the trivial cohomology class which includes
states of ghost numbers $0$ and $(+1)$, all four nontrivial classes
are in one-to-one correspondence with each of the four basis states
$|\pm\pm>$ in the ghost sector. {\sl A priori\/}, each of these
nontrivial classes could also be put into one-to-one correspondence with
the physical states spanned by $|n_0>$ that define the correct 
quantization of this system as constructed in Sect.\ref{Sect3}, up to
the product with the $\hat{p}_\lambda$-eigenstate $|p_\lambda=0>$. However, 
when considering the time evolution of BRST invariant states generated by 
the Schr\"odinger equation to be discussed below, it turns out that only the 
BRST nontrivial cohomology class of minimal ghost number $(-1)$ always
possesses the correct time dependent dynamics irrespective of the choice of
gauge fixing fermion $\Psi$, leading back precisely to the characterization
of physical states given in (\ref{eq:physicalBRST}) as being only those BRST
invariant states that are of minimal ghost number. This is a general
feature of Hamiltonian BRST quantization whatever the gauge invariant system
being considered. On the other hand, even though in the present system it
might appear that such a one-to-one correspondence with Dirac's
identification of physical states could possibly also apply for the other
nontrivial cohomology classes at ghost numbers 0 and $(+1)$, whether their
correct time evolution is recovered as well is again a matter dependent on
the choice of gauge fixing function $\Psi$, and is generally not realized,
even for an admissible gauge fixing.

Furthermore, as we shall see shortly, the presence of the extended
phase space degrees of freedom, leading to the $|p_\lambda=0>\otimes|-->$ 
component of the physical states, implies an ill defined normalization
of BRST invariant matrix elements of the BRST invariant quantum evolution
operator, namely {\sl in fine\/}, of the physical states themselves.
For instance, although the physical component $|n_0>$ is properly
normalized to unity, that of the BRST invariant physical state 
(\ref{eq:physicalBRST}) itself, whether $\psi_{--,n_0}(t)$ is a pure phase 
factor or not, is ill defined since the inner product of the state 
$|p_\lambda=0>$ with itself is the $\delta$-function $\delta(0)$ while that 
of the state $|-->$ with itself vanishes, leading to an undetermination of 
the form $0\cdot\delta(0)$ for the inner product of the physical states in
(\ref{eq:physicalBRST}).

The same issue also arises for BRST invariant matrix elements of the
BRST invariant quantum evolution operator of the system. Given the class
of gauge fixing functions of the form (\ref{eq:Psi}), the corresponding
BRST invariant quantum Hamiltonian is
\begin{equation}
\hat{H}_{\rm eff}=\frac{i}{\hbar}\{\hat{\Psi},\hat{Q}_B\}=
\hat{\lambda}\left[\hbar\omega(a^\dagger a+\frac{1}{2})-\Lambda\right]+
F(\hat{\lambda})\hat{p}_\lambda-i\hbar\hat{c}^1\hat{b}_1F'(\hat{\lambda})
-i\hbar\hat{c}^1\hat{b}_2\ ,
\end{equation}
for which the Schr\"odinger equation over the extended space of quantum
states reads
\begin{equation}
i\hbar\frac{d}{dt}|\psi,t>=\hat{H}_{\rm eff}|\psi,t>\ .
\end{equation}
Applied to the nontrivial BRST cohomology classes in 
(\ref{eq:genBRST}), one finds that for this class of gauge fixing
choices only $\psi_{--,n_0}(t)$ and $\psi_{-+,n_0}(t)$ are necessarily
time independent as is required for actual gauge invariant states in Dirac's
approach, whereas $\psi_{+-,n_0}(t)$ never is, while $\psi_{++,0}(t)$ would
be provided only $F(\lambda)$ were to be constant, which still excludes a
large class of admissible gauge fixing choices. Hence, it is only
the cohomology class of minimal ghost number constructed in
(\ref{eq:physicalBRST}) that always also solves the Schr\"odinger equation
in a manner consistent with the actual identification of physical states in
Dirac's quantization, namely in the present case with a time independent
component $\psi_{--,n_0}$. In spite of the normalization problems just
mentioned, there thus appears a one-to-one correspondence between the
physical states $|n_0>\psi_{--,n_0}$ of Dirac's quantization as discussed
in Sect.\ref{Sect3} and the BRST invariant physical states of minimal ghost
number constructed here. The physical content should  thus be equivalent in
both approaches, provided however that time evolution in the BRST approach
be also consistent with that in Dirac's approach.

To understand how gauge fixing and Gribov problems may interfere with
such an equivalence precisely at that level, let us now consider
the BRST invariant quantum evolution operator associated to a given
choice of gauge fixing function $\Psi$, namely the operator
\begin{equation}
\hat{U}_{\rm BRST}(t_2,t_1)=e^{-i/\hbar(t_2-t_1)\hat{H}_{\rm eff}}=
e^{(t_2-t_1)\{\hat{\Psi},\hat{Q}_B\}/\hbar^2}\ .
\end{equation}
Naively, matrix elements of this operator between BRST invariant
states should be independent of the gauge fixing function $\Psi$,
since the BRST charge is nilpotent, $\hat{Q}^2_B=0$. Indeed, given
two states $|\psi_1>$ and $|\psi_2>$ that are annihilated by $\hat{Q}_B$,
it would appear\cite{BFV} that through a direct expansion of the exponential,
one has simply
\begin{equation}
<\psi_2|\hat{U}_{\rm BRST}(t_2,t_1)|\psi_1>=<\psi_2|\psi_1>\ ,
\end{equation}
thus apparently establishing that BRST invariant matrix elements
of $\hat{U}_{\rm BRST}(t_2,t_1)$ are totally independent of the
choice of gauge fixing function $\Psi$, namely the usual
statement of the Fradkin-Vilkovisky theorem\cite{BFV}.
However as we have observed, the matrix element on the r.h.s. of this 
relation is ill defined for nontrivial BRST cohomology classes (or at best
it vanishes identically, which is certainly not what should be expected from
matrix elements representing a quantum system with actual physical content), 
so that the l.h.s. requires further specification as to its 
explicit evaluation, thereby possibly 
implying that even though such matrix elements are indeed BRST invariant, 
and thus gauge invariant as well, they may
not be totally independent of the gauge fixing function $\Psi$. Indeed,
this is what actually happens, and once again, it is only for an admissible
gauge fixing, namely such that each of the distinct gauge orbits of
the dynamical system is effectively singled out once and only once,
that the correct physics is represented through the above
matrix elements\cite{Gov1,Gov2}. Nevertheless in the general case, the above
matrix elements do depend on the choice of gauge fixing but actually then
only through the gauge equivalence class to which that gauge fixing belongs,
these gauge equivalence classes being defined through the covering of the
space of gauge orbits which is induced by the gauge fixing. In fact, this is
the actual content of the Fradkin-Vilkovisky theorem\cite{Gov1,Gov2}.

As a first illustration of these features within the present example,
let us consider the zero ghost number BRST invariant states that are 
associated to the BRST invariant boundary conditions (\ref{eq:BRSTbc}), namely
\begin{equation}
\begin{array}{r c l}
|\psi_i>&=&|q_i>\otimes|p_\lambda=0>\otimes|-+>=
\sum_{n=0}^\infty|n>\otimes|p_\lambda=0>\otimes|-+><n|q_i>\ \ \ ,\\
|\psi_f>&=&|q_f>\otimes|p_\lambda=0>\otimes|-+>=
\sum_{n=0}^\infty|n>\otimes|p_\lambda=0>\otimes|-+><n|q_f>\ ,
\end{array}
\end{equation}
where $|q>$ denotes the $\hat{q}$-eigenstate basis normalized such
that\footnote{The matrix elements $<n|q>$ are thus
the harmonic oscillator energy eigenstate configuration space wave
functions in the case of our specific example.}
$<q|q'>=\delta(q-q')$. However, a direct evaluation of the matrix elements
\begin{equation}
<n;p_\lambda=0;-+|\hat{U}_{\rm BRST}(t_2,t_1)|m;p_\lambda=0;-+>\ ,
\end{equation}
leads to an undetermination of the form $0\cdot\delta(0)$, as explained
previously. To avoid this ambiguity, one may momentarily relax the
requirement of BRST invariance of the external states\footnote{An
alternative approach which maintains explicit BRST invariance at all
stages will be presented in Ref.\cite{GS}, confirming once again the
discussion developed here.} and consider rather the matrix elements
\begin{equation}
<n;p_{\lambda,2};-+|\hat{U}_{\rm BRST}(t_2,t_1)|m;p_{\lambda,1};-+>\ .
\label{eq:MA}
\end{equation}
Being no longer BRST invariant, this expression may now acquire a well
defined value which, however, may also depend on the choice for $\Psi$.
Hence even in the limit $p_{\lambda,1},p_{\lambda,2}\rightarrow 0$ of
BRST invariant external states, which might lead to a
distribution-like expression, it could be that the final result obtained 
through this form of evaluation through a continuity procedure of an
otherwise ill defined expression retains some dependency on $\Psi$,
albeit a gauge invariant one through the induced covering of the space
of gauge orbits. As we shall see, this is indeed the case, and as a
general result is the actual content of the Fradkin-Vilkovisky
theorem\cite{Gov1,Gov2}.

The evaluation of the matrix elements (\ref{eq:MA}) is best performed
through a path integral representation. The $(\lambda,p_\lambda)$
and ghost sectors being common to all one-dimensional reparametrization
invariant systems, and thus common to that of the relativistic scalar
particle in a Minkowski spacetime, one may borrow the result
of the path integral evaluation over these degrees of freedom directly
from that latter system\cite{Gov1,Gov2}. With the following
choice of sign for the ghost matrix elements $<--|++>=i=-<-+|+->$,
and given the class
of gauge fixing functions $\Psi$ in (\ref{eq:Psi}), one then finds the
exact and explicit result,
\begin{equation}
\lim_{p_{\lambda,1},p_{\lambda,2}\rightarrow 0}
<n;p_{\lambda,2};-+|\hat{U}_{\rm BRST}(t_2,t_1)|m;p_{\lambda,1};-+>=
i\delta_{n,m}\int_{\cal D}\frac{d\gamma}{2\pi\hbar}\,
e^{-i\gamma/\hbar(\hbar\omega(n+1/2)-\Lambda)}\ ,
\label{eq:BRSTPI}
\end{equation}
where the real Teichm\"uller parameter\cite{Gov1,Gov2}
\begin{equation}
\gamma=\int_{t_1}^{t_2}dt\,\lambda(t)\ ,
\end{equation}
the Lagrange multiplier $\lambda(t)$ and the domain of integration ${\cal D}$ 
are constructed as follows. Given the gauge fixing condition
\begin{equation}
\dot{\lambda}=F(\lambda)
\end{equation}
associated to (\ref{eq:Psi}), and an arbitrary integration constant
$\lambda_f=\lambda(t_f)$, consider the corresponding solution 
$\lambda(t;\lambda_f)$ to this equation as well as its Teichm\"uller
parameter $\gamma(\lambda_f)$. As the integration 
constant $\lambda_f$ covers the interval $]-\infty,+\infty[$ once with that 
orientation, the Teichm\"uller parameter $\gamma(\lambda_f)$ covers a certain 
oriented domain ${\cal D}$ in the real line. This is the domain over which 
the final integral in (\ref{eq:BRSTPI}) must be performed, including its 
orientation. This is also how all the extended phase space degrees of freedom 
contribute to the final matrix element, leading to a dependency on the Lagrange 
multiplier only through its gauge invariant Teichm\"uller 
parameter\footnote{The Teichm\"uller parameter
is indeed invariant under all local reparametrizations $\epsilon(t)$
consistent with the chosen boundary conditions at $t=t_{i,f}$ for the
matter degrees of freedom, namely $\epsilon(t_{i,f})=0$.}.

Even though the final expression (\ref{eq:BRSTPI}) is indeed manifestly 
BRST and gauge invariant, it does depend on the choice of gauge fixing
function $\Psi$ or $F(\lambda)$ in as far as the domain of integration
${\cal D}$ depends on that choice, namely precisely through the covering of
the space of gauge orbits that is induced by the choice of gauge fixing
function $F(\lambda)$. An admissible gauge fixing is such
that the domain ${\cal D}$ coincides exactly once with the real
line\cite{Gov1,Gov2},
which is the case for example for $F(\lambda)=\alpha\lambda+\beta$.
However, if one considers for instance $F(\lambda)=\alpha\lambda^3$
with $\alpha>0$ and $(t_2-t_1)>0$, the domain ${\cal D}$ is such that
the Teichm\"uller parameter $\gamma$ ranges from 
$-\sqrt{2(t_2-t_1)/\alpha}$ to $\sqrt{2(t_2-t_1)/\alpha}$,
clearly displaying the dependency of (\ref{eq:BRSTPI}) on the parameter
$\alpha$ defining the gauge fixing condition, albeit in a gauge invariant
manner. Nevertheless, in the limit
$\alpha\rightarrow 0$, one recovers the domain associated to an
admissible gauge fixing, as is indeed then also the choice $F(\lambda)=0$.
Hence, contrary to what is often stated, the BRST invariant matrix
elements of the BRST invariant quantum evolution operator are not totally
independent of the gauge fixing condition, but depend on it only in as far
as the induced covering of the space of gauge orbits is dependent on the
choice of gauge fixing. For a gauge fixing procedure
suffering Gribov ambiguities, such as the choices $F(\lambda)=\alpha\lambda^3$
or $F(\lambda)=\alpha\lambda^2+\beta\lambda+\delta$ when $\alpha\ne 0$,
one does not represent the same quantum dynamics as that of the
original system, in spite of the fact that the description remains
manifestly gauge invariant irrespective of the gauge fixing choice
and procedure. Gauge invariance is not all there is to gauge invariant
systems!

Assuming now an admissible gauge fixing choice to have been made,
in which case the integral in ({\ref{eq:BRSTPI}) is over the entire
real line, one thus finally has the following evaluation of the
relevant BRST invariant matrix elements which represent the quantum
propagator for physical states within the BFV-BRST Hamiltonian
formalism,
\begin{equation}
<n;p_\lambda=0;-+|\hat{U}_{\rm BRST}(t_2,t_1)|m;p_\lambda=0;-+>=
i\delta_{n,m}\delta\left(\hbar\omega(n+\frac{1}{2})-\Lambda\right)\ .
\label{eq:final}
\end{equation}
Hence once again, but clearly only provided the gauge fixing procedure
is free of any Gribov ambiguity, one recovers the requirement of a
quantized value of the cosmological constant which must belong to the
energy spectrum of the matter sector, namely in the present instance
for some specific positive integer $n_0$,
\begin{equation}
\Lambda=\hbar\omega(n_0+\frac{1}{2})\ ,
\end{equation}
in which case the physical subspace is also confined to the states
spanned by $|n_0>$ in the matter sector.

As a matter of fact, provided a specific choice for $F(\lambda)$ is made
at the outset, it is also possible to circumvent the path integral
evaluation of the relevant matrix element. For example given the admissible
gauge fixing with $F(\lambda)=0$, as well as the following non-BRST invariant
external states
\begin{equation}
|\psi_i>=\int_{-\infty}^\infty dp_\lambda\,
|n;p_\lambda;-+>\,\psi_i(p_\lambda)\ \ ,\ \ 
|\psi_f>=\int_{-\infty}^\infty dp_\lambda\,
|m;p_\lambda;-+>\,\psi_f(p_\lambda)\ ,
\end{equation}
an explicit calculation of the matrix element
$<\psi_f|\hat{U}_{\rm BRST}(t_2,t_1)|\psi_i>$ is readily performed,
with the result
\begin{equation}
<\psi_f|e^{-i/\hbar(t_2-t_1)\hat{H}_{\rm eff}}|\psi_i>=
i\delta_{n,m}(t_2-t_1)\int_{-\infty}^\infty d\lambda\,
\tilde{\psi}^*_f(\lambda)\,
e^{-i/\hbar\lambda(t_2-t_1)(\hbar\omega(n+1/2)-\Lambda)}\,
\tilde{\psi}_i(\lambda)\ ,
\end{equation}
where
\begin{equation}
\tilde{\psi}_{i,f}(\lambda)=\int_{-\infty}^\infty\frac{dp_\lambda}
{\sqrt{2\pi\hbar}}\,e^{i/\hbar\,\lambda p_\lambda}\,\psi_{i,f}(p_\lambda)\ .
\end{equation}
In the limit of BRST invariant external states, namely $\delta$-peaked
in the conjugate momentum $p_\lambda$ with 
$\psi_{i,f}(p_\lambda)=\delta(p_\lambda)$ and thus non-normalizable,
one recovers the above expression (\ref{eq:BRSTPI}) in the case
of an admissible gauge fixing condition, namely with ${\cal D}$ being
the entire real line,
\begin{equation}
<n;p_\lambda=0;-+|\hat{U}_{\rm BRST}(t_2,t_1)|m;p_\lambda=0;-+>=
i\delta_{n,m}\int_{-\infty}^\infty \frac{d\gamma}{2\pi\hbar}\,
e^{-i/\hbar\gamma(\hbar\omega(n+1/2)-\Lambda)}\ ,
\end{equation}
and thus the same final expression and conclusion as in (\ref{eq:final}).
Nevertheless, one sees that whatever the approach, the evaluation
of such BRST invariant matrix elements, associated to the quantum propagator
of physical states, requires first an evaluation for non-BRST invariant 
external states, leading therefore to the possibility of a dependency of
gauge invariant quantities on the gauge fixing conditions, albeit in a gauge
invariant manner. In fact, it is possible to circumvent the ill defined
singular products $0\cdot\delta(0)$ stemming from the non-normalizable
character of the $\hat{p}_\lambda$ eigenstates in still another manner,
that also explicitly preserves manifest BRST invariance at all stages and 
does not require to consider $\hat{U}_{\rm BRST}(t_2,t_1)$
matrix elements for non-BRST invariant external states, and still reach
exactly the identical conclusion\cite{GS}.
Any gauge fixing leads to a dynamics defined over the space of gauge orbits
of the system, hence to a gauge invariant dynamics, but which of these gauge 
orbits are included and with which multiplicity is dependent on the
choice of gauge fixing procedure. The correct physics and dynamics is
represented only for an admissible gauge fixing procedure, namely one
which selects once and only once each of the gauge orbits and thus induces
an admissible covering of the space of gauge orbits. And it is only for such
an admissible gauge fixing that the correct quantization of the
cosmological constant is reproduced within BRST quantization.

As a final remark, note that when the cosmological constant takes
a quantized valued within the energy spectrum of the matter sector,
$\Lambda=E_{n_0}$, the relevant BRST invariant matrix elements are given in 
terms of the $\delta(E_n-\Lambda)$ distribution with a singular value for the
physical sector $n=n_0$, yet another manifestation
of the lack of normalizability of physical states
within the BRST quantization approach which is due to the
extended phase space degrees of freedom. This situation is to be contrasted
with the identification of physical states within Dirac's approach, in
which everything is so much more straightforward and transparent when
expressed in terms of the physical projector, including the dynamical time
evolution of physical quantum states.

\section{Conclusions}
\label{Sect6}

Inspired by a simple example discussed in Ref.\cite{Klauder}, this note has
discussed the coupling of quantum gravity in one dimension to an arbitrary
quantum mechanical interacting matter sector. One of the main conclusions
of this analysis is that, as a very general result, and even though 
one-dimensional gravity is quantum dynamics free, the cosmological
constant in such systems needs to take a value that belongs to the energy
spectrum of the quantum matter sector, and thus to be quantized, while
physical states are then also confined to the subspace of matter quantum
states whose energy coincides with the cosmological constant value.
These are certainly very intriguing conclusions, which call for their possible
extension to higher dimensional theories of quantum gravity coupled to
interacting quantum matter, hopefully eventually shedding some new light onto
the cosmological constant problem.

The quantization of these one-dimensional matter coupled quantum gravity 
systems is most efficient and simple based on the physical projector 
approach\cite{Proj,Klauder}, lending itself to the immediate identification
of the gauge invariant physical sector and its quantum dynamics,
and of the quantization requirement on the 
cosmological constant. Such a circumstance enables one to contrast the 
efficacy of that approach to that of other methods developed over the years 
towards the quantization of gauge invariant, and more generally,
of constrained dynamics.

As has been explicitly illustrated, Faddeev's reduced phase space
approach\cite{Faddeev} simply fails for one-dimensional
reparametrization invariant systems, for reasons that may be traced back to 
the appearance of Gribov ambiguities whatever the gauge fixing conditions that
might be introduced\cite{Gov2,Klauder}. Even though the resulting
formulation remains gauge invariant, it is not independent of the gauge
fixing conditions\cite{Gov2}, and thus does not necessarily reproduce the same
quantum physics as that of the original system, as is indeed always the case
in the presence of Gribov  problems. Given that most approaches towards
quantum gravity rely one way or another on Faddeev's gauge fixing procedure,
the fact that it already fails in one dimension because of Gribov problems and
thus cannot capture the then necessary quantization condition on the
cosmological constant leaves totally open the possibility that such a
quantization is also required in higher dimensions.

The Hamiltonian BRST invariant quantization method\cite{BFV} has also been
applied to these classes of systems, to explicitly illustrate once
again the dependency of the resulting quantum dynamics on the gauge fixing
condition\cite{Gov1,Gov2}. It is only for an admissible gauge fixing procedure
that BRST invariant physical states as well as the actual BRST invariant
quantum dynamics are in one-to-one correspondence with the physical states
and their correct quantum dynamics as readily identified within Dirac's
approach through the physical projector, and that the requirement of a
quantized cosmological constant is recovered. Nevertheless, the introduction
of extended degrees of freedom, among which the ghost sector required to
balance the gauge variant variables, implies that BRST invariant physical
states correspond to non-normalizable quantum states, a technical difficulty
which is avoided altogether within Dirac's approach and the physical
projector.

These general classes of one-dimensional matter coupled quantum gravity
models clearly illustrate the various advantages of the physical
projector approach, in which no gauge fixing procedure whatsoever is
required and which thus from the outset avoids\cite{Gov3} having to address
the difficult and subtle issues of Gribov problems which plague essentially
any gauge fixing conditions being used for systems of physical interest.
Beginning with the intriguing question regarding the possible
quantized value of the cosmological constant and the confinement of physical
states to the subspace of matter quantum states whose energy coincides
with cosmological constant value, the exploration of
the efficacy and the new insights that the physical projector
may bring to constrained dynamics thus appears to be of potential great
interest to the understanding of the gauge invariant theories on which
our present description of the physical world and its fundamental
interactions is based\cite{QG1,Gov4}. Such an understanding might also bring
us closer to determining whether the suggestion made in the opening
paragraph of the Introduction has any chance of being physically
meaningful.

\section*{Acknowledgements}

It is my great pleasure to dedicate this note to John R. Klauder
on the occasion of his 70$^{\rm th}$ birthday, as a tribute to
his many scientific insights and his warm friendship. I also wish to
thank the Stellenbosch Institute for Advanced Study (STIAS) and its Director,
Prof. Bernard Lategan, for financial support making possible
my participation in the STIAS hosted Workshop on String Theory and Quantum
Gravity (February 4-22, 2002), a most enjoyable and rewarding
experience indeed, during which this work was completed. My thanks also go
out to Prof. Hendrik Geyer, organizer of the Workshop, and Prof. Frederik
Scholtz, Head of the Department of Physics at the University of
Stellenbosch, for their wonderful hospitality,
and for fruitful scientific discussions.

\end{document}